\documentclass[
    amsmath,
    amssymb,
    superscriptaddress,
    aps,
    prx,
    reprint,
    floatfix,
]{revtex4-2}

\usepackage{graphicx}
\usepackage{dcolumn}
\usepackage{hyperref}
\usepackage{glossaries}
\usepackage[version=3]{mhchem}
\usepackage[per-mode=symbol]{siunitx}
\usepackage[most]{tcolorbox}
\usepackage{xcolor}
\usepackage{booktabs}

\hypersetup{
    unicode,
    colorlinks=true,
    urlcolor=black,
    linkcolor=black,
    citecolor=black
}

\renewcommand{\vec}[1]{\ensuremath\boldsymbol{#1}}

\newacronym{acf}{ACF}{autocorrelation function}
\newacronym{dft}{DFT}{density-functional theory}
\newacronym{md}{MD}{molecular dynamics}
\newacronym{nep}{NEP}{neuroevolution potential}
\newacronym{npt}{NPT}{isothermal-isobaric}
\newacronym{qnep}{qNEP}{charge-aware neuroevolution potential}

\newcommand{\gpumd}{\textsc{gpumd}}
\newcommand{\ase}{\textsc{ASE}}
\newcommand{\calorine}{\textsc{Calorine}}
\newcommand{\vasp}{\textsc{vasp}}

\makeatletter
\let\oldtheequation\theequation
\renewcommand\tagform@[1]{\maketag@@@{\ignorespaces#1\unskip\@@italiccorr}}
\renewcommand\theequation{(\oldtheequation)}
\makeatother

\DeclareSIUnit\angstrom{\text{Å}}
\DeclareSIUnit\site{\text{site}}
\DeclareSIUnit\atom{\text{atom}}

\begin{document}

\title{Anomalous polar chains in halide perovskites}

\newcommand{\addchalmersphy}{Department of Physics and Astronomy, Chalmers University of Technology, SE-412 96 Gothenburg, Sweden}

\newcommand{\dur}{
    Centre for Materials Physics, Durham University, South Road, Durham, DH1 3LE, United Kingdom
}

\author{Erik Fransson}
\email{erikfr@chalmers.se}
\affiliation{\addchalmersphy}

\author{Apinya Ngoipala}
\affiliation{\addchalmersphy}

\author{Priyanka Deswal}
\affiliation{\addchalmersphy}

\author{Ethan Berger}
\affiliation{\addchalmersphy}

\author{Nicholas C. Bristowe}
\affiliation{\dur}

\author{Paul Erhart}
\affiliation{\addchalmersphy}

\author{Julia Wiktor}
\email{julia.wiktor@chalmers.se}
\affiliation{\addchalmersphy}

\date{\today}

\begin{abstract}
Local polar distortions, their collective ordering, and their field-driven reversal have been studied extensively in ferroelectric perovskites for decades.
Here we identify an unusual behavior in the ferroelectric halide perovskite \ce{CsGeBr3} using large-scale molecular dynamics with first-principles-trained interatomic potentials.
In its nominally cubic phase the Ge off-centering forms one-dimensional polar chains that extend over at least tens of nanometers and relax on nanosecond timescales, four orders of magnitude slower than in \ce{BaTiO3} and \ce{PbTiO3}, with \ce{CsSnBr3} in between.
The cubic \ce{CsGeBr3} phase is a dynamically disordered state in which oppositely polarized chains flip throughout the crystal, and a long-lived one-dimensional local polarization coexists with zero net polarization.
A simple three-dimensional perovskite thus hosts long-ranged one-dimensional polar correlations with remarkably slow dynamics.
In the ferroelectric phase the same chains set a one-dimensional switching mechanism, in which the polarization reverses chain by chain.
\end{abstract}

\maketitle

\section{Introduction}

Ferroelectrics possess a spontaneous polarization that can be switched between symmetry-equivalent states by an electric field \cite{Valasek1921,Lines1977}, underpinning applications in memory, sensing, and actuation \cite{Scott2007,Rabe2007}.
Oxide perovskites such as \ce{BaTiO3} and \ce{PbTiO3} are well-studied examples, yet their atomic-scale behavior, particularly that of \ce{BaTiO3}, continues to invite discussion.
A central question is how collective lattice vibrations, described by soft-mode theory \cite{Cochran1960,Cohen1992}, and the ordering of local off-centered displacements, described by order--disorder models \cite{Comes1968,Chaves1976}, combine to produce the ferroelectric transition \cite{Bruce1980,Zalar2003,Senn2016}.
In cubic \ce{BaTiO3}, anisotropic displacement correlations have been interpreted through both one-dimensional chains \cite{Comes1968,Lambert1969,Senn2016} and strongly anharmonic soft-mode dynamics \cite{Huller1969,Pasciak2018}, a question revisited recently with large-scale simulations and real-space imaging \cite{Gigli2022,Zhang2025a}.
Calculations have also identified favorable parallel alignment along chains together with antipolar coupling between neighboring chains \cite{Zhang2006}.
Connecting these local arrangements to their spatial extent and relaxation dynamics remains important for a coherent microscopic picture of ferroelectricity.

Halide perovskites provide less extensively studied counterparts to these oxide ferroelectrics.
Much attention has focused on Pb-based compounds, where B-site off-centering is comparatively weak, while the stereochemically active $ns^2$ lone pair \cite{Orgel1959,Walsh2011,Fabini2020} produces pronounced local distortions in Sn- and especially Ge-based systems \cite{Stoumpos2015,Fabini2016,Laurita2017,HyltonFarrington2026}.
The polar rhombohedral structure of \ce{CsGeCl3} was reported already in 1965 \cite{Christensen1965}, and the structural transitions of \ce{CsGeX3} were characterized two decades later and interpreted in terms of order--disorder behavior \cite{Thiele1987}.
More recent experiments have established ferroelectricity and examined local coordination \cite{Zhang2022,Hooper2022}, while effective-Hamiltonian models \cite{Kashikar2024,Townsend2024,Kashikar2026}, first-principles calculations \cite{Zhang2026a} and first-principles molecular dynamics \cite{HyltonFarrington2026} have revealed a strong order--disorder character, competing polar and antipolar arrangements, and slow Ge reorientation.
However, connecting these findings across the spatial and temporal scales of collective ordering remains challenging.
In particular, the extent, dynamics, and role of correlated displacements in polarization reversal remain less well understood than in oxide prototypes.

Here, we investigate \ce{CsGeBr3} using large-scale \gls{md} simulations with machine-learned interatomic potentials trained on \gls{dft} data.
These potentials allow us to approach first-principles accuracy on the length and time scales accessible to effective-Hamiltonian models \cite{Zhong1994}, while retaining an explicit description of all atoms.
Through comparison with \ce{BaTiO3}, \ce{PbTiO3}, and \ce{CsSnBr3}, this enables a detailed, quantitatively grounded microscopic picture connecting local distortions, collective correlations, and ferroelectric switching.
We find pronounced longitudinal correlations in cubic \ce{CsGeBr3}, accompanied by short-ranged transverse antipolar tendencies and slow displacement relaxation.
We further show how chain-wise organization supports effectively one-dimensional polarization-reversal pathways, connecting the unusual local dynamics to the material's ferroelectric switching behavior.
Instead of a domain wall sweeping through the crystal, the polarization reverses through asynchronous flips of individual one-dimensional polar chains.
This reveals how low-dimensional switching can emerge within a chemically simple, three-dimensional bulk perovskite.

\section{Methods}

\subsection{Machine-learned potentials}

We used two types of machine-learned potentials, \gls{nep} and \gls{qnep}, implemented in \gpumd{} \cite{FanZenZha21,Fan22,FanWanYin22,XuBuPan25}.
Unless stated otherwise, results were obtained with \gls{qnep}, which includes environment-dependent atomic charges and long-range electrostatic interactions \cite{Fan2026qnep}; a comparison with the corresponding \gls{nep} model for \ce{CsGeBr3} is given in \autoref{sfig:lattice_vs_T_nep_vs_qnep} and \autoref{sfig:snapshot_dx_nep_vs_qnep}.

\textbf{\ce{BaTiO3}.}
For \ce{BaTiO3}, we employ the previously developed \gls{qnep} model of Ref.~\cite{Fan2026qnep}, which describes its construction and validation.

\textbf{\ce{PbTiO3}.}
For \ce{PbTiO3}, new \gls{nep} and \gls{qnep} models were trained in this work using reference \gls{dft} data obtained with the r$^2$SCAN exchange--correlation functional \cite{Furness2020} as implemented in the \vasp{} code \cite{KreFur96a}.
We used the same plane-wave cutoff (\qty{520}{\eV}) and $k$-point spacing (\qty{0.18}{\per\angstrom}) as used in the training of the \ce{BaTiO3} potential \cite{Fan2026qnep}.
The training dataset includes primitive cells with various structural distortions, supercells with random displacements, and predominantly snapshots from \gls{md} simulations at temperatures ranging from \qty{0}{\kelvin} to \qty{1000}{\kelvin}.

\textbf{\ce{CsGeBr3} and \ce{CsSnBr3}.}
For the two halide perovskites we employ separate \gls{nep} and \gls{qnep} models, each trained on \gls{dft} data obtained with the r$^2$SCAN50 hybrid functional \cite{Furness2020,Bursch2022}, which admixes \qty{50}{\percent} exact exchange, as implemented in the all-electron electronic-structure code \textsc{fhi-aims} \cite{blum2009ab}.
The reference calculations used the \emph{intermediate} default numerical atom-centered basis set \cite{blum2009ab}.
This functional has been shown to be accurate for phase transitions in Ge perovskites, predicting transition temperatures close to experimentally observed values \cite{Priyanka-paper,Kashikar2026}.
Both models additionally include the screened nuclear repulsion potential of Ziegler, Biersack, and Littmark \cite{Ziegler1985} as a short-range baseline, as implemented in \gpumd{}.
The \ce{CsGeBr3} models build on those developed in Ref.~\cite{Priyanka-paper}, with additional training structures added to specifically cover the high-temperature cubic phase.
The validation of the potentials against \gls{dft} is summarized in the Supplementary Information (\autoref{sfig:parity} and \autoref{sfig:phonons_pes}).

\subsection{Molecular dynamics simulations}
All structure handling and the interface to \gpumd{} were managed through \ase{}~\cite{Larsen2017} and \calorine{}~\cite{calorine}.
Production \gls{md} simulations were performed with \gpumd{}~\cite{FanWanYin22,XuBuPan25} using $24 \times 24 \times 24$ supercells containing \num{69120} atoms, and always at zero pressure.
A timestep of \qty{1}{\femto\second} was used for the oxides and \qty{5}{\femto\second} for the halides.

Heating simulations were run in the \gls{npt} ensemble using stochastic cell rescaling \cite{Bernetti2020}, as implemented in \gpumd{}.
Temperature was increased continuously from \qty{1}{\kelvin} to above each transition over \qty{20}{\nano\second} for the oxides and \qty{100}{\nano\second} for the halides.
Phase transitions were identified from the lattice parameters and angles (\autoref{sfig:lattice_vs_T}).
The cubic phases were analyzed at temperatures approximately \qtyrange{15}{80}{\kelvin} above the respective transition temperatures.

Static and dynamic correlations were sampled in the $NVT$ and $NVE$ ensembles, respectively, following \qty{5}{\nano\second} of equilibration.
These simulations used fixed cells with the average lattice parameters at zero pressure obtained from \gls{npt} runs.
Equilibration was tested extensively for \ce{CsGeBr3} because of its slow dynamics (\autoref{sfig:ge_structure_equilibration} and \autoref{sfig:ge_energy_equilibration}).
Temporal correlations were evaluated from short trajectories of \qtyrange{50}{100}{\pico\second}, supplemented by longer trajectories of \qty{2.5}{\nano\second} for \ce{CsSnBr3} and \qty{100}{\nano\second} for \ce{CsGeBr3}.
Spatial correlations were sampled in the $NVT$ ensemble over \qty{1}{\nano\second} for the oxides and \qty{100}{\nano\second} for the halides.

Polarization--electric-field (P--E) loops were computed using the Born effective charges predicted by the \gls{qnep} model, following Ref.~\cite{Fan2026qnep}.
The loops were run in the \gls{npt} ensemble with the field along [111], at \qty{100}{\kelvin} for \ce{BaTiO3} and \qty{300}{\kelvin} for \ce{CsGeBr3}.
After equilibration without field, the field was swept continuously from zero to $+E_\mathrm{max}$, to $-E_\mathrm{max}$, back to $+E_\mathrm{max}$ and to zero over \num{15000000} steps, with $E_\mathrm{max} = \qty{10}{\milli\volt\per\angstrom}$ for \ce{BaTiO3} and \qty{60}{\milli\volt\per\angstrom} for \ce{CsGeBr3}, corresponding to sweeps of \qty{15}{\nano\second} and \qty{75}{\nano\second} and similar ramp rates of \num{4.0} and \qty{4.8}{\milli\volt\per\angstrom\per\nano\second}.
The polarization was estimated as $\mathbf{P}=V^{-1}\sum_i \mathbf{Z}_i^{}\mathbf{u}_i$, where $V$ is the volume, $\mathbf{Z}_i^{}$ are fixed Born effective charges evaluated with \gls{qnep} at the relaxed ground-state structure, and $\mathbf{u}_i$ are the atomic displacements from an ideal cubic reference.

\subsection{B-site off-centering}
The B-site off-centering was quantified by the displacement
\begin{align}
    \vec{d} = \vec{r}_\mathrm{B} - \frac{1}{6}\sum_{i=1}^{6} \vec{r}_{\mathrm{X},i},
\end{align}
where $\vec{r}_\mathrm{B}$ is the position of the B-site cation (Ti, Ge, or Sn) and the second term is the geometric center of the surrounding octahedron, defined by its six neighboring anions X (O or Br).

\subsection{Correlations}

Spatial and temporal correlations were evaluated from the B-site displacements $\vec{d}(\mathbf{R})$ in saved frames of the cubic-phase $NVT$ and $NVE$ trajectories, respectively.
Here, $\mathbf{R}$ labels a site on the reference simple-cubic B sublattice and $a$ is its lattice parameter.

For separations of $n$ lattice spacings along a cubic axis, we define the dimensionless displacement correlation
\begin{align}
    C_{\alpha\beta}(n) = \frac{\left\langle d_\alpha(\mathbf{R})\, d_\alpha(\mathbf{R} + n a \hat{\mathbf{e}}_\beta) \right\rangle}{\left\langle d_\alpha^2 \right\rangle},
\end{align}
where $\alpha$ denotes the Cartesian displacement component, $\hat{\mathbf{e}}_\beta$ is the unit vector along the separation direction, and $\langle\cdot\rangle$ denotes an average over all B sites and saved frames.
In short, $C_{xy}(n)$ is the correlation of B-site off-centering displacement in the x-direction between B-site neighbors along the y-direction.
In the cubic phase, symmetry leaves two distinct functions: the longitudinal correlation $C_L(n)$ for $\alpha=\beta$ and the transverse correlation $C_T(n)$ for $\alpha\neq\beta$.
We average the numerator and denominator over the three equivalent longitudinal pairs or six equivalent transverse pairs before taking their ratio.

The temporal behavior of the off-centering is characterized by the \gls{acf} of the displacement vector,
\begin{align}
    C(\tau) = \frac{\left\langle \vec{d}(t) \cdot \vec{d}(t + \tau) \right\rangle}{\left\langle \left| \vec{d} \right|^2 \right\rangle},
\end{align}
where the average runs over all B sites and all time origins $t$, and the normalization  gives $C(0) = 1$.
This single-site \gls{acf} measures the persistence of the displacement, including changes in both its magnitude and direction.

\section{Results and discussion}

\subsection{Local distortion}

\begin{figure*}[ht]
    \centering
    \includegraphics{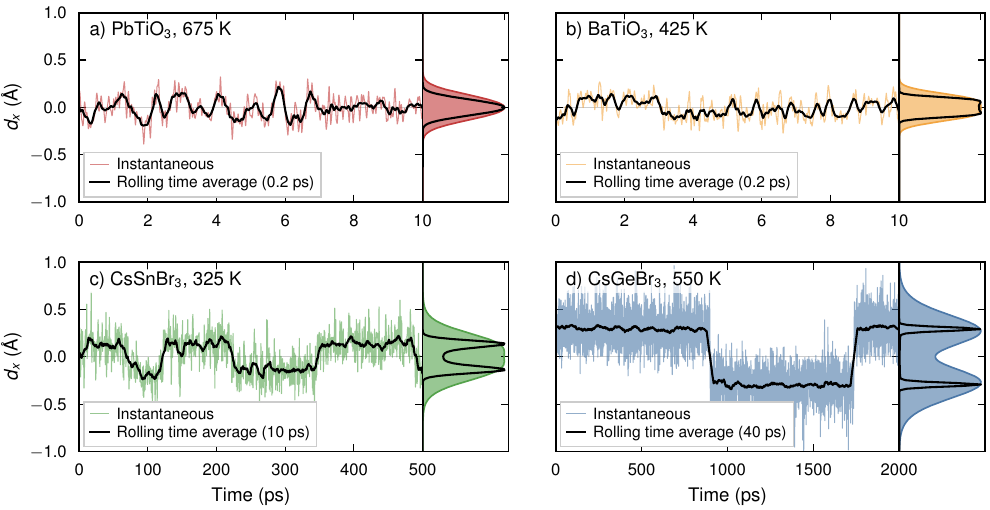}
    \caption{
        \textbf{Time evolution of the B-site off-centering displacement.}
        Off-centering component $d_x(t)$ of one selected B site for a) \ce{PbTiO3} at \qty{675}{\kelvin}, b) \ce{BaTiO3} at \qty{425}{\kelvin}, c) \ce{CsSnBr3} at \qty{325}{\kelvin}, and d) \ce{CsGeBr3} at \qty{550}{\kelvin}, in the $24 \times 24 \times 24$ supercells.
        The $x$ component is shown as one representative; by cubic symmetry the $y$ and $z$ components behave equivalently (see \autoref{sfig:snapshot_CGB_components}).
        Black lines show a centered rolling time average of $d_x(t)$, with windows of \qty{0.2}{\pico\second}, \qty{0.2}{\pico\second}, \qty{10}{\pico\second}, and \qty{40}{\pico\second} in a) to d), respectively.
        The narrow side panels show the distribution of the instantaneous $d_x$, collected over all B sites and all times, sharing the $y$ axis with the time trace; the horizontal axis is the probability density.
        Black histograms show the time-averaged displacement, with their scale matched to the instantaneous distributions.
    }
    \label{fig:single_site_dx}
\end{figure*}

The cubic average structure conceals distinct local displacement dynamics (\autoref{fig:single_site_dx}).
A useful simplified picture is motion in a local multiwell landscape, with off-centered minima along the eight $\langle111\rangle$ directions and an unstable high-symmetry position at the center \cite{Comes1968,Pirc2004}.
We write a Cartesian displacement component as
\begin{align}
    d(t) = d_{\mathrm{static}}(t) + d_{\mathrm{vib}}(t).
    \label{eq:d_decomposition}
\end{align}
Here, $d_{\mathrm{static}}(t)$ is the corresponding component of the occupied well's position relative to the high-symmetry site, and $d_{\mathrm{vib}}(t)$ is the displacement about that minimum.
The static contribution changes as the atom hops between wells, while the vibrational contribution describes motion within the occupied well.
When hopping is slow compared with vibration, a time average over an intermediate window approximates $d_{\mathrm{static}}(t)$.

In the order--disorder limit, the transition primarily removes the directional order of persistent off-centered configurations; in the displacive limit, it is associated with the softening of a collective vibration \cite{Rubtsov2000}.
The rapid Ti fluctuations in \autoref{fig:single_site_dx} are consistent with the prominent displacive soft-mode response of \ce{PbTiO3}, although experiments and recent simulations also identify substantial order--disorder character in its transition \cite{Sicron1994,Xie2025}.
For \ce{BaTiO3}, atomistic simulations and NMR support mixed displacive and order--disorder transition character \cite{Zhong1995,Zalar2003,Qi2016}.
Recent real-space imaging directly reveals persistent local Ti displacements across the ferroelectric--paraelectric transition, providing further evidence for its order--disorder component \cite{Zhang2025a}.
The halides show clearer hopping between off-centered configurations, which is very slow in \ce{CsGeBr3}, consistent with the order--disorder interpretation of its structural transition \cite{Thiele1987}.
In \ce{CsSnBr3}, vibrational broadening can obscure the off-centered populations in the instantaneous distribution, while time averaging reveals their bimodality (\autoref{sfig:displacement_histograms}).
This supports local order--disorder dynamics, consistent with experimental evidence for dynamic Sn off-centering, without assigning the same mechanism to its octahedral-tilting transitions \cite{Fabini2016}.

The single-site dynamics already reveal unusually persistent local distortions in \ce{CsGeBr3}, but do not establish whether these evolve independently or through collective motion.
To connect this local picture to the behavior of the crystal, we next examine the spatial and temporal correlations of the B-site displacements.

\subsection{Correlations}

\begin{figure*}[ht]
    \centering
    \includegraphics{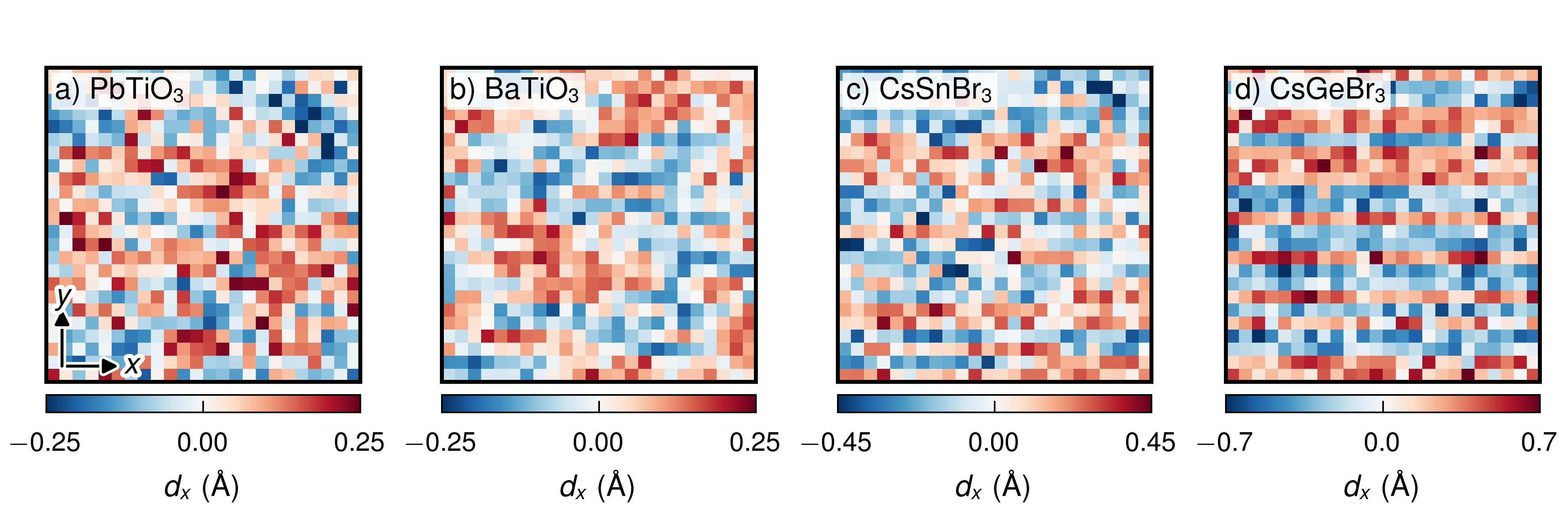}
    \caption{
        \textbf{Instantaneous snapshots of the B-site off-centering displacement.}
        Off-centering component $d_x$ of every B site in a single instantaneous $(001)$ layer of the $24 \times 24 \times 24$ supercell for a) \ce{PbTiO3} at \qty{675}{\kelvin}, b) \ce{BaTiO3} at \qty{425}{\kelvin}, c) \ce{CsSnBr3} at \qty{325}{\kelvin}, and d) \ce{CsGeBr3} at \qty{550}{\kelvin}.
        Only the $x$ component is shown; the choice is arbitrary, and the $y$ and $z$ components show the same general pattern along their respective directions.
        Note the different color scales; the two oxides share one scale.
        Time-averaged snapshots of the same configurations are shown in \autoref{sfig:snapshot_time_average}.
    }
    \label{fig:snapshots}
\end{figure*}

\begin{figure*}[ht]
    \centering
    \includegraphics{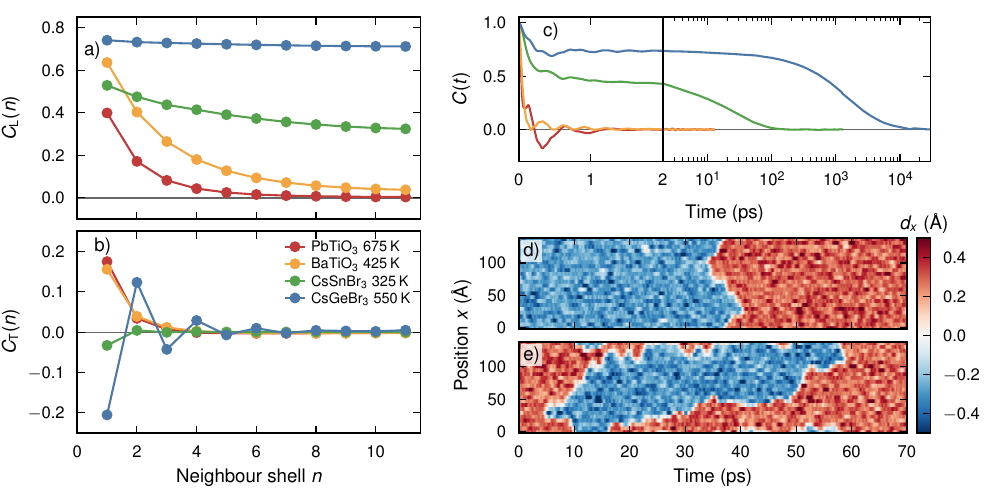}
    \caption{
        \textbf{Spatial and temporal correlations of the B-site off-centering in the cubic phase.}
        a) Longitudinal correlation $C_\mathrm{L}(n)$ and b) transverse correlation $C_\mathrm{T}(n)$ as a function of neighbor shell $n$, and c) autocorrelation $C(t)$ of the off-centering, for \ce{PbTiO3} at \qty{675}{\kelvin}, \ce{BaTiO3} at \qty{425}{\kelvin}, \ce{CsSnBr3} at \qty{325}{\kelvin} and \ce{CsGeBr3} at \qty{550}{\kelvin} in $24 \times 24 \times 24$ supercells.
        The shell $n = 0$, where both correlations equal one by normalization, is omitted.
        The time axis in c) is linear below \qty{2}{\pico\second} and logarithmic above.
        Panels a)-c) are averages over all B sites and all frames.
        d,e) Off-centering component $d_x$ along two $x$-oriented chains of a single \ce{CsGeBr3} trajectory at \qty{550}{\kelvin} and zero applied field, as a function of time and of position along the chain, averaged over \qty{1}{\pico\second} per site.
        These are two selected chains, not ensemble averages.
    }
    \label{fig:joint_correlations}
\end{figure*}

The instantaneous snapshots in \autoref{fig:snapshots} give a first impression of how the off-centering is organized in space.
In the two oxides, $d_x$ varies from site to site with only faint patches of common distortions, and no preferred direction is apparent in a single frame.
In \ce{CsSnBr3} faint streaks running along $x$ can be discerned, and in \ce{CsGeBr3} the pattern is unmistakable: the $x$ component of the off-centering forms stripes that span the entire layer along $x$, while neighboring stripes along $y$ tend to alternate in sign.
The displacement is thus correlated along its own direction, forming polar one-dimensional chains of parallel off-centering, whereas correlations perpendicular to the displacement appear weak.
Time averaging sharpens the same pattern (\autoref{sfig:snapshot_time_average}), and the same stripes appear in \ce{MAGeI3} (\autoref{sfig:snapshot_instant}) \cite{MAGeSn-paper}, which indicates that the chain formation follows from the Ge off-centering itself rather than from the A-site cation or the halide.

The correlation functions in \autoref{fig:joint_correlations}a,b) make this quantitative.
The longitudinal correlation $C_\mathrm{L}(n)$ is positive in all four systems, but its range differs by more than an order of magnitude.
In \ce{PbTiO3} it vanishes within three to four lattice spacings, consistent with the sub-nanometer dipole correlations reported for its paraelectric phase \cite{Xie2025}, and in \ce{BaTiO3} within roughly ten, in line with chain lengths of about 10 to 60 unit cells estimated for its cubic phase from diffuse scattering and simulations \cite{Comes1968,Lambert1969,Pasciak2018}.
In \ce{CsSnBr3} the correlation decays slowly and is still about 0.3 at the largest separation accessible in the supercell.
In \ce{CsGeBr3} it barely decays at all, indicating that the one-dimensional chains remain almost perfectly correlated throughout the entire supercell.
This persists far above the transition, well over \qty{100}{\kelvin} into the cubic phase (\autoref{sfig:correlations}).

The transverse correlation $C_\mathrm{T}(n)$ is short-ranged in all four systems, but its sign distinguishes them.
The oxides show a weak positive coupling between neighboring chains that decays within a few shells.
In \ce{CsSnBr3} the nearest-neighbor transverse correlation is slightly negative and beyond that essentially zero, so the chains are stacked without noticeable order.
Interestingly, in \ce{CsGeBr3}, $C_\mathrm{T}(n)$ alternates in sign, from about $-0.2$ at $n = 1$ to $+0.1$ at $n = 2$, and the oscillation dies out after four to five shells.
Neighboring chains thus prefer to be antiparallel, which is the alternating stripe pattern seen in \autoref{fig:snapshots}d), but this antipolar order extends only over a few chains.
The preference for antiparallel neighboring chains is what dipole--dipole interactions between neighboring chains would favor \cite{LuttingerTisza1946,Zhong1995}.
The same stacking is obtained with the \gls{nep} model without explicit long-range electrostatics (\autoref{sfig:correlations_nep_vs_qnep}), so it does not require explicit long-range electrostatics.
It also connects to effective-Hamiltonian simulations, in which the same cubic phase was described as a dynamic antipolar state made up of antipolar nanoregions \cite{Kashikar2024,Kashikar2026}.
Our results support this antipolar coupling, but show that the long-ranged order in the cubic phase is the polar correlation along the chains, while the antipolar coupling between chains is weak and survives only locally, over a few chains.
Why this antipolar tendency appears in \ce{CsGeBr3} but is much weaker in \ce{CsSnBr3}, whose ground state is itself antipolar \cite{Fabini2024}, is a question we do not resolve here.

The temporal correlations in \autoref{fig:joint_correlations}c) span an even wider range between the systems.
In both oxides the \gls{acf} decays to zero within about \qty{1}{\pico\second}, and for \ce{PbTiO3} the damped oscillation of an anharmonic soft mode is clearly visible.
In the halides the \gls{acf} drops within the first picosecond to a plateau before decaying on a much longer timescale.
The short-timescale behavior corresponds to the vibrational part, $d_\mathrm{vib}$, while the plateau measures the static part, $d_\mathrm{static}$, which only decays once the atoms hop between wells.
For \ce{CsSnBr3} this second decay takes many tens of picoseconds, and for \ce{CsGeBr3} nanoseconds, with the \gls{acf} reaching zero only after about \qty{10}{\nano\second}.
This is consistent with the single-site time traces in \autoref{fig:single_site_dx}.
The persistence of the off-centering thus increases by at least four orders of magnitude from the oxides to \ce{CsGeBr3} (see also \autoref{sfig:acfs}).

The long chains and the slow relaxation are two aspects of the same motion, which is illustrated by the space--time maps in \autoref{fig:joint_correlations}d,e).
Along a chain the off-centering is uniform over the full length of about \qty{140}{\angstrom} for tens of picoseconds, and it reverses spontaneously in a collective fashion.
In \autoref{fig:joint_correlations}d) the entire chain switches sign within several picoseconds, whereas in \autoref{fig:joint_correlations}e) a reversed segment nucleates, grows to cover most of the chain, and then collapses back to the original orientation.
These processes take place throughout the system at once, independently for the $x$, $y$ and $z$ chains, giving rise to a rich polar chain-flipping dynamics on the nanosecond timescale.
Reversal of a full chain thus proceeds by nucleation and growth of a reversed segment along the chain, the kink dynamics expected for a classical one-dimensional double-well chain \cite{Krumhansl1975,Buttiker1979,Hinzke2000}.
The same behavior is found for larger supercells and over longer time windows (\autoref{sfig:kymograph_sizes}).

The present simulations do not resolve the intrinsic longitudinal correlation length in \ce{CsGeBr3}.
In larger supercells, individual chains remain uniformly polarized over more than \qty{400}{\angstrom} (\autoref{sfig:kymograph_sizes}), but their slow relaxation further complicates convergence.
Because longer chains also flip more slowly, the length and time scales are coupled through the supercell size, which also underlies the strong size dependence of the simulated transition temperature of \ce{CsGeBr3} reported in Ref.~\cite{Priyanka-paper}.
In larger supercells we furthermore observe that the antipolar stacking of the chains, which is short-ranged in the $24 \times 24 \times 24$ cells analyzed above, can sometimes develop into long-range order over about \qty{100}{\nano\second}, with chains of one orientation forming an antipolar pattern across the entire supercell (\autoref{sfig:size_ordering}).
Whether this points to a metastable or even stable antipolar state of bulk \ce{CsGeBr3}, or is a consequence of the finite periodic supercells, cannot be decided from the present simulations.
Resolving these correlations, and the possibility of such a state, would require for example single-crystal scattering measurements, as discussed further in the Supplementary Information.
Diffuse-scattering measurements could help constrain this length scale through the widths of the narrow reciprocal-space features associated with these correlations, which appear as lines in the simulated planar cuts of $S(\vec q)$ shown in the Supplementary Information (\autoref{sfig:sq} and \autoref{sfig:sq_L10}).

\subsection{Ferroelectric switching}

\begin{figure*}[ht]
    \centering
    \includegraphics{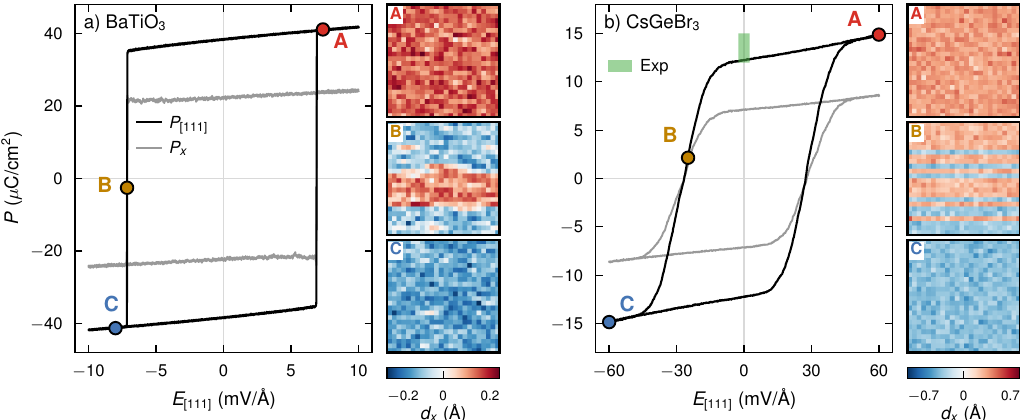}
    \caption{
        \textbf{Field-driven polarization switching.}
        Polarization projected along [111] and its $x$ component as a function of applied [111] field, for a) \ce{BaTiO3} at \qty{100}{\kelvin} and b) \ce{CsGeBr3} at \qty{300}{\kelvin}.
        Insets show the B-site off-centering component $d_x$ in one $(001)$ layer at the marked points A--C, taken in a) from an equivalent run of the same protocol.
        Each inset averages five consecutive frames separated by \qty{15}{\pico\second} in b) and a single frame in a).
        Note the different $d_x$ color scales in a) and b).
        The green bar (Exp) in b) marks the experimental room-temperature spontaneous polarization of \qtyrange{12}{15}{\micro\coulomb\per\centi\metre\squared} measured for \ce{CsGeBr3} \cite{Zhang2022}.
    }
    \label{fig:pe_loop}
\end{figure*}

The chain correlations in \ce{CsGeBr3} are not only a feature of the cubic phase.
They also shape how the polarization reverses in the ferroelectric phase, which we examine with field-driven switching simulations of \ce{CsGeBr3} at room temperature, using \ce{BaTiO3} as a reference (\autoref{fig:pe_loop}).
Both materials are run in their rhombohedral phases, \ce{BaTiO3} at \qty{100}{\kelvin} and \ce{CsGeBr3} at \qty{300}{\kelvin}, with the field applied along the polar [111] direction, so that both loops connect the same pair of $\langle111\rangle$ states.

\ce{BaTiO3} shows a very clear square loop in which the polarization reverses in almost a single step once the coercive field, the field at which the polarization vanishes, is reached.
The snapshot at the switching point (\autoref{fig:pe_loop}a, inset B) shows extended reversed and unreversed regions separated by sharp domain walls.

\ce{CsGeBr3} behaves differently, showing a much more rounded loop as the reversal is spread over a wide range of fields (\autoref{fig:pe_loop}b).
At intermediate fields some chains have already switched while their neighbors have not, so the system consists of a mixture of reversed and unreversed chains.
The polarization thus reverses chain by chain, asynchronously, rather than through a domain that sweeps through the crystal.
Space--time maps of neighboring chains make this explicit: in \ce{CsGeBr3} each chain flips as a unit at its own time, separated from its neighbors by nanoseconds, whereas in \ce{BaTiO3} all chains flip together (\autoref{sfig:kymograph_switching}).
After a step reversal of a constant field of \qty{30}{\milli\volt\per\angstrom}, close to the coercive field of the loop, the switched fraction follows the Kolmogorov--Avrami--Ishibashi model with an Avrami exponent of $n \approx 1$ (\autoref{sfig:avrami}), which is inconsistent with the growth of two- or three-dimensional domains and consistent with chains that reverse individually \cite{Ishibashi1971}.
The same one-dimensional nucleation-and-growth process that flips chains in the cubic phase thus appears to govern switching in the ferroelectric phase, now driven by the field rather than by thermal fluctuations.
The remanent polarization, about \qty{12}{\micro\coulomb\per\centi\metre\squared}, lies at the lower end of the experimental range for \ce{CsGeBr3} \cite{Zhang2022}.

With both materials in their rhombohedral phases, which for our \ce{BaTiO3} model requires a temperature as low as \qty{100}{\kelvin} (\autoref{sfig:lattice_vs_T}), the coercive field of \ce{CsGeBr3} is roughly four times that of \ce{BaTiO3}.
This is only a qualitative measure, since simulated coercive fields depend on sweep rate and system size \cite{Beckman2009,Kashikar2026}, but the difference reflects an important distinction between the two switching mechanisms.
In \ce{BaTiO3} all chains reverse together within picoseconds once a reversed domain has formed, whereas in \ce{CsGeBr3} a reversed segment grows readily along its own chain but does not trigger the reversal of neighboring chains, which flip nanoseconds apart (\autoref{sfig:kymograph_switching}).
Each chain therefore requires its own nucleation event, consistent with the Avrami exponent of $n \approx 1$, and this raises the apparent coercive field at a finite sweep rate even though growth along each chain is easy.
In addition, the polar ground state of \ce{CsGeBr3} lies about six times deeper below the cubic structure than that of \ce{BaTiO3}, which likely also contributes to its slower switching.

A one-dimensional switching pathway in a simple three-dimensional perovskite is uncommon and sets the Ge perovskite apart from better-known and more extensively studied ferroelectrics.
Low-dimensional polar behavior is usually tied to low-dimensional structures, whereas here it emerges in a bulk three-dimensional material from the strong correlations of the Ge off-centering alone.
A related exception is \ce{HfO2}, where flat polar phonon bands make atomic-scale polar layers independently switchable \cite{Lee2020}, and the transverse polar phonon branches of \ce{CsGeBr3} are similarly flat (Supplementary Information).
The chain coupling therefore offers a handle for tuning the switching behavior, for example by B-site alloying with Pb or Sn.
The gradual, chain-by-chain reversal could also give access to well-defined intermediate polarization states, of the kind sought for analog and multilevel memory, and might in principle allow memristor-like behavior down to the level of single polar chains.

\section{Conclusions}
By combining machine-learned interatomic potentials with large-scale \gls{md} simulations, we have shown that the Sn and Ge off-centering in their cubic phases is organized in one-dimensional polar chains, especially in \ce{CsGeBr3}, where the chains show remarkable extent and persistence.
In \ce{CsGeBr3} the chains remain correlated across the entire supercell and the off-centering survives for nanoseconds, orders of magnitude longer than in \ce{BaTiO3} and \ce{PbTiO3} and far longer than in \ce{CsSnBr3}, while neighboring chains prefer to stack antiparallel.
The cubic phase is therefore best pictured as a dynamically disordered state in which extended, oppositely polarized chains flip continuously throughout the crystal, reminiscent of the polar nanoregions of relaxor ferroelectrics but one-dimensional and far longer ranged.
The slow relaxation and the long chains are two aspects of the same phenomenon, since a chain reverses through nucleation and growth of a reversed segment along its length, which is the dynamics of a classical one-dimensional double-well chain.
This mechanism carries over to field-driven switching in the ferroelectric phase, where the polarization reverses chain by chain with each chain requiring its own nucleation event, and this, together with the deeper polar wells, raises the coercive field above that of \ce{BaTiO3}.
Diffuse scattering should reveal the chains directly as sharp planar features in reciprocal space, and B-site alloying offers a route to tune the chain coupling and with it the switching behavior.

\section*{Data availability}
The data supporting this study are openly available via Zenodo at \url{https://zenodo.org/records/22007823}.

\section*{Acknowledgements}
We acknowledge funding from the Swedish Foundation for Strategic Research through a Future Research Leader programme (FFL21-0129), the European Research Council (ERC Starting Grant no. 101162195), and the Knut and Alice Wallenberg Foundation (Nos.~2023.0032 and 2024.0042).
The computations were enabled by resources provided by the National Academic Infrastructure for Supercomputing in Sweden (NAISS), partially funded by the Swedish Research Council through grant agreement no. 2022-06725, and by the Berzelius resource provided by the Knut and Alice Wallenberg Foundation at the National Supercomputer Centre (NSC).

The authors used AI tools (Claude, Anthropic, and ChatGPT, OpenAI) during the preparation of this manuscript for assistance with writing and language improvements.
All scientific content, interpretation, and final wording were reviewed and approved by the authors.

\bibliography{references}

\end{document}